\documentclass[sigconf]{acmart}
\usepackage{multirow}
\usepackage{subcaption}
\usepackage{url}

\begin{document}

\title{Cross-Task Benchmarking and Evaluation of General-Purpose and Code-Specific Large Language Models}

\author{Gunjan Das}
\authornote{Work done during the internship at Bosch Research and Technology Centre, India}
\affiliation{%
   \institution{National Institute of Technology Karnataka}
   \country{India}
}
\email{dasgunjan2004@gmail.com}

\author{Paheli Bhattacharya, Rishabh Gupta}
\affiliation{%
  \institution{Bosch Research and Technology Centre}
  \country{India}
  }
\email{{paheli.bhattacharya, gupta.rishabh}@in.bosch.com}





\begin{abstract}
  Large Language Models (LLMs) have revolutionized both general natural language processing and domain-specific applications such as code synthesis, legal reasoning, and finance. However, while prior studies have explored individual model capabilities, a systematic cross-domain comparison that unifies linguistic, reasoning, and code understanding abilities remains underexplored. In this work, we present a comprehensive evaluation of five general-purpose and three code-specific state-of-the-art LLMs across six diverse benchmarks encompassing linguistic competence, mathematical reasoning, and trustworthiness. Additionally, we analyze model behavior on the CoNaLa dataset for code explanation, comparing natural language and code-specialized LLMs. Our findings reveal that models optimized for code (e.g., CodeLLaMA variants) exhibit strong reasoning and syntactic precision, that even for non-coding tasks can show measurable performance gains, in contrast to general-purpose models like Mistral-7B and Llama-3-8B.

\end{abstract}



\maketitle

\section{Introduction}
Large Language Models (LLMs) have made remarkable progress across a wide range of domains, from general natural language processing to specialized tasks such as code synthesis and explanation~\cite{bhattacharya-gupta-2025-selective}. As these models continue to evolve, systematically evaluating their performance across diverse applications has become increasingly critical. In particular, understanding how domain-optimized models, such as those designed for code generation, perform on traditional natural language tasks remains an open research question.

Several survey studies have highlighted the diversity of available LLMs, each tailored to specific application domains [9, 15]. However, there remains a notable gap in comparative understanding of these models across heterogeneous benchmarks spanning linguistic competence, commonsense and mathematical reasoning, and trustworthiness. Such a holistic view is essential for informed model selection, enabling practitioners to choose models that best align with task requirements. While evaluation metrics for individual benchmarks are publicly available through leaderboard repositories, a unified and systematic consolidation of these results has been lacking. Our work addresses this gap by bringing together benchmark outcomes under a single analytical framework, offering empirical guidance on model suitability for different use-case priorities-for instance, identifying which model performs best when trustworthiness is more important than linguistic fluency.

Recent years have witnessed the rise of domain-specific LLMs such as CodeLlama [17] for programming, ChatLaw [6] for legal reasoning, and BloombergGPT [21] for financial analytics. However, the cross-domain performance of such models-for example, assessing how a code-specialized model like CodeLlama performs on mathematical or reasoning benchmarks-remains underexplored. Addressing this gap can reveal how domain specialization influences general reasoning and language understanding capabilities.

In this paper, we present a comprehensive comparative analysis of five general-purpose and five code-specific LLMs across six diverse benchmarks evaluating different aspects of language generation-namely, linguistic competence, commonsense reasoning, mathematical reasoning, and trustworthiness. We further extend our evaluation to the CoNaLa dataset [22], focusing on the task of code explanation, where both model categories are assessed using fine-grained metrics such as BLEU, ROUGE, and CodeBERTScore. This unified assessment yields several noteworthy findings-for instance, while Llama-3-8B [7] exhibits strong linguistic performance, it falls short in trustworthiness, whereas CodeLlama-34B demonstrates broader reasoning strengths.

To the best of our knowledge, this is the first study to consolidate and contrast such a wide range of LLMs across both natural language and code-oriented tasks, thereby providing an integrated perspective that informs model selection for diverse applications.

\section{Methodology}
In this paper, we survey two families of LLMs --  (i) General purpose LLMs like Llama-3-8B that are trained on linguistic datasets in general and not explicitly have any domain-specific capabilities. ~(ii) Code specific LLMs like CodeLlama-13b~\cite{roziere2023code} that expertise on Code related tasks. 
We perform a comparative analysis of these LLMs through the lenses of different natural language generation metrics available in the form of benchmarks. We first describe the LLMs and then the benchmarks studied in this paper.

\subsection{Selection of LLMs}
In this work, we methodically selected a diverse array of state-of-the-art LLMs to assess their performance across a spectrum of general language tasks. The selection encompasses models that represent a wide range of architectural paradigms, scales, and training objectives, thus offering a comprehensive survey of the current state-of-the-art in the field. \\

\noindent
The models included from the family of general purpose LLMs are:

$\bullet$ \textbf{Llama2-7B, 13B~\cite{touvron2023llama} and Llama3-8B~\cite{dubey2024llama}}: These models, from Meta's LLaMA series, are distinguished by their robustness in natural language understanding and generation, providing strong baselines for a variety of tasks.

$\bullet$ \textbf{Vicuna-7B~\cite{zheng2024judging}}: Vicuna is widely acknowledged for its fine-tuning towards enhanced performance in dialogue and conversational AI tasks, making it an essential inclusion for evaluating performance in interactive settings.

$\bullet$ \textbf{Mistral-7B~\cite{jiang2023mistral}}: Mistral introduces a new generation of lightweight yet potent models, emphasizing efficiency while maintaining competitive performance across both general and specialized tasks.

$\bullet$ \textbf{DeepSeek-R1:1.5B~\cite{guo2025deepseek}}: A lightweight yet powerful reasoning model, designed for efficiency and enhanced logical consistency across diverse domains. Its compact architecture enables rapid inference while maintaining strong performance on analytical and problem-solving tasks, making it ideal for studying reasoning capabilities in smaller-scale LLMs.\\

\noindent
The models included from the family of code specific LLMs are:

$\bullet$ \textbf{CodeLlama-34B} and \textbf{CodeLlama-13B-Instruct~\cite{roziere2023code}}: These models are specifically optimized for code-centric tasks, providing critical insights into the adaptability of LLMs to the unique challenges of code generation.

$\bullet$ \textbf{StarCoder~\cite{li2023starcoder}}: This model also specialized in code, recognized for its ability to generate syntactically correct and contextually appropriate code, which is crucial for evaluating LLM performance in programming tasks.

\subsection{Benchmark Dataset and Evaluation Metrics}
In this section, we describe the different benchmark datasets on which LLMs are evaluated. Each metric provides a distinct perspective on the model's capabilities, facilitating a comprehensive analysis of their effectiveness. The following metrics are utilized in our comparative study:

\subsection{Natural Language (NL) Metrics}

Here we briefly describe the datasets that translate into metrics for evaluating the natural language capabilities of LLMs. 

$\bullet$ \textbf{Multitask Model Language Understanding (MMLU)~\cite{mmlu}} evaluates a model’s performance across a wide range of language tasks, providing an aggregate score that reflects its ability to handle diverse linguistic challenges. This metric is crucial for assessing the versatility and overall proficiency of LLMs in NLP.

$\bullet$ \textbf{ARC~\cite{arc}} metric evaluates a model's accuracy on the ARC Challenge, which features complex multiple-choice science questions designed to test reasoning and scientific knowledge. It is crucial for assessing models' deep understanding and logical reasoning in scientific contexts, highlighting their proficiency in the technical domains.

$\bullet$ \textbf{HellaSwag~\cite{hellaswag}} is a benchmark for evaluating a model's commonsense reasoning by requiring it to complete sentences in story contexts. It tests the model's ability to use everyday knowledge and generate plausible, coherent continuations, assessing its handling of contextual information and commonsense reasoning in natural language scenarios.


$\bullet$ \textbf{Winogrande~\cite{winogrande}} is a large-scale dataset that assesses commonsense reasoning through sentence completion tasks involving nuanced understanding and ambiguity resolution. It tests a model’s ability to apply commonsense knowledge effectively, making it crucial for evaluating performance in complex reasoning scenarios beyond basic comprehension.


$\bullet$ \textbf{TruthfulQA~\cite{truthfulqa}} evaluates a model's ability to generate accurate and truthful responses, focusing on its integrity and reliability. This metric is essential for applications where avoiding misinformation and ensuring trustworthiness are critical.


$\bullet$ \textbf{Grade-School Math 8K (GSM8K)~\cite{gsm}} evaluates a model's quantitative reasoning through grade-school level math problems, focusing on arithmetic and basic problem-solving tasks. It is crucial for assessing a model’s ability to handle numerical reasoning and mathematical operations.

\subsection{Code Explanation Metrics}
To evaluate an LLMs ability to understand and generate natural language descriptions of code snippets, we use the \textbf{CoNaLa} \textbf{(Code/Natural Language Challenge)}~\cite{conala} dataset. Given an input code snippet, the task is to generate intents/explanation that the code is achieving. This focuses on semantic code understanding, intent modeling and text generation. The dataset consists of $1,666$ Python code and paired natural language annotations. The average length of a code snippet and its corresponding explanation are both 14 tokens on average~\cite{bhattacharya-gupta-2025-selective, bhattacharya2023exploring}. Below are some examples from the dataset :\\

\noindent
\textbf{Code}: \texttt{sum = lambda x, y: x + y} \\
\textbf{Explanation}: This defines a lambda function to sum two numbers.

\vspace{1mm}
\noindent
\textbf{Code}: \texttt{re.findall(r'\textbackslash d+', text)} \\
\textbf{Explanation}: Finds all numbers in a string using regex. 

\vspace{2mm}
\noindent
The generated explanations are then compared to the reference annotations provided in the dataset. We use the following standard metrics that capture various aspects such as lexical overlap and semantic similarity using the following standard evaluation metrics for the task~\cite{bhattacharya-gupta-2025-selective, geng, segment}.

\noindent
(i)~\textbf{Token-based}: These metrics collectively quantify surface-level similarity, lexical accuracy, and fluency between generated and reference texts. \textbf{BLEU}~\cite{bleu} evaluates n-gram precision using a weighted geometric mean across multiple n-gram lengths, emphasizing exact lexical overlap. \textbf{METEOR}~\cite{meteor} extends this by incorporating stemming, synonymy, and word order, balancing precision and recall through a harmonic mean to better reflect semantic similarity. \textbf{ROUGE}~\cite{rouge} metrics adopt a recall-oriented perspective: ROUGE-1 measures unigram recall for content relevance, ROUGE-2 captures bigram-level fluency through short phrase overlap, and ROUGE-L leverages the longest common subsequence (LCS) to assess overall sequence similarity and structural coherence.

\noindent
(ii)~\textbf{Semantics-based}: We use this measure to assess the semantic similarity between the model generated explanation ($m$) and the ground truth explanation ($g$). We project both $m$ and $g$ in a continuous embedding space, $ \overrightarrow{e_m}$ and $ \overrightarrow{e_g}$ respectively using the pretrained CodeBERT~\cite{codebert} model. We then take a cosine similarity between the embeddings $cosine( \overrightarrow{e_m},  \overrightarrow{e_g})$ to get the score. We refer to this metric as \textbf{CodeBERT} in the rest of the paper.

\subsection{Experimental Design}
In our study, we evaluated the performance of various LLMs using a diverse set of metrics, including ARC, MMLU, TruthfulQA, HellaSwag, Winogrande and GSM8K. These metrics collectively offer a comprehensive assessment of each model's capabilities across both general natural language tasks.

To benchmark these LLMs, we leveraged the Hugging Face Open LLM Leaderboard\footnote{\url{https://huggingface.co/spaces/open-llm-leaderboard/open_llm_leaderboard}}, which provides valuable insights into general language generation metrics. However, it should be noted that this leaderboard does not include metrics specifically designed for code generation.

To address this limitation, we extended our evaluation by submitting the models to the leaderboard, focusing on their performance in generation tasks, particularly those involving code generation. This approach allowed us to gather detailed data on each model's ability to generate accurate and contextually relevant outputs. By synthesizing these results, we aim to offer a nuanced understanding of the strengths and weaknesses of different LLMs in both general-purpose and code-specific contexts.

\vspace{-3mm}
\begin{figure}[!htbp]
\centering
\caption{Performance of general purpose and code specific LLMs on natural language (a, b) and code metrics (c, d).}
\label{fig:combined_metrics}
\vspace{-3mm}
\begin{subfigure}{0.49\textwidth}
\centering
\caption{General purpose LLMs on NL metrics}
\includegraphics[width=\textwidth]{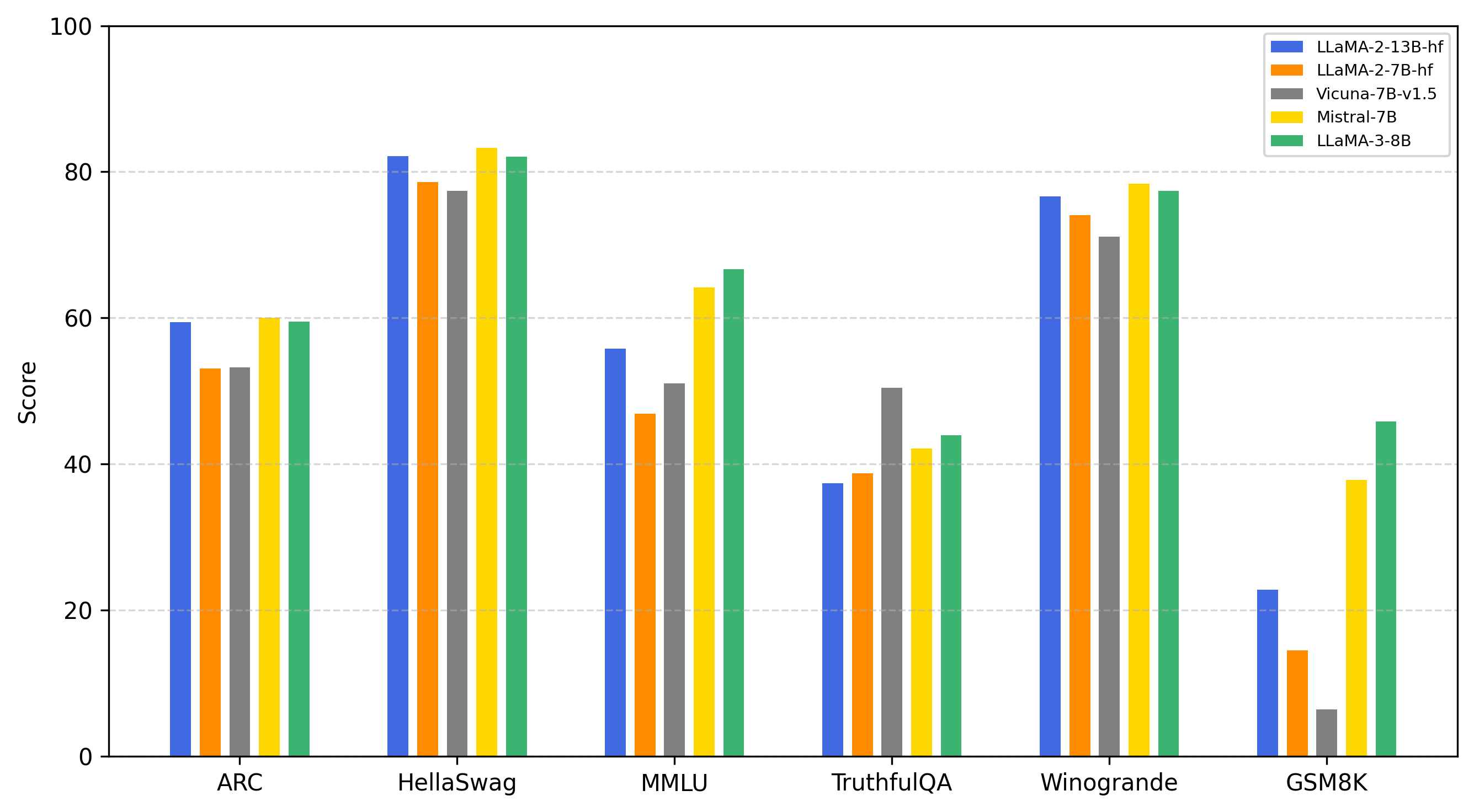}
\label{fig:gen_llm_metrics}
\end{subfigure}
\vspace{-8mm}

\begin{subfigure}{0.49\textwidth}
\centering
\caption{General Purpose LLMs on Code Explanation metrics}
\includegraphics[width=\textwidth]{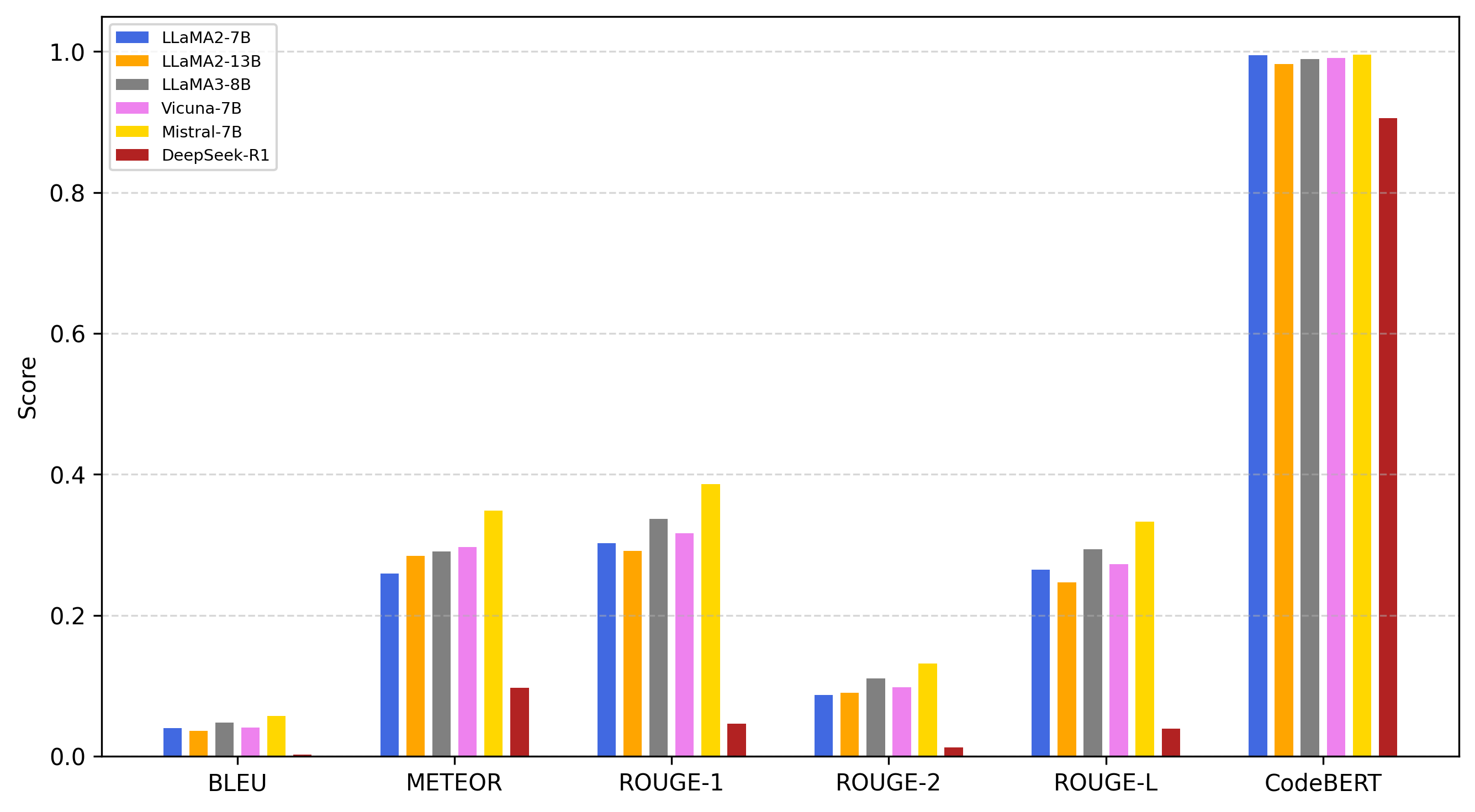}
\label{fig:gen_llm_metrics_code}
\end{subfigure}
\vspace{-8mm}

\begin{subfigure}{0.49\textwidth}
\centering
\caption{Code specific LLMs on NL metrics}
\includegraphics[width=\textwidth]{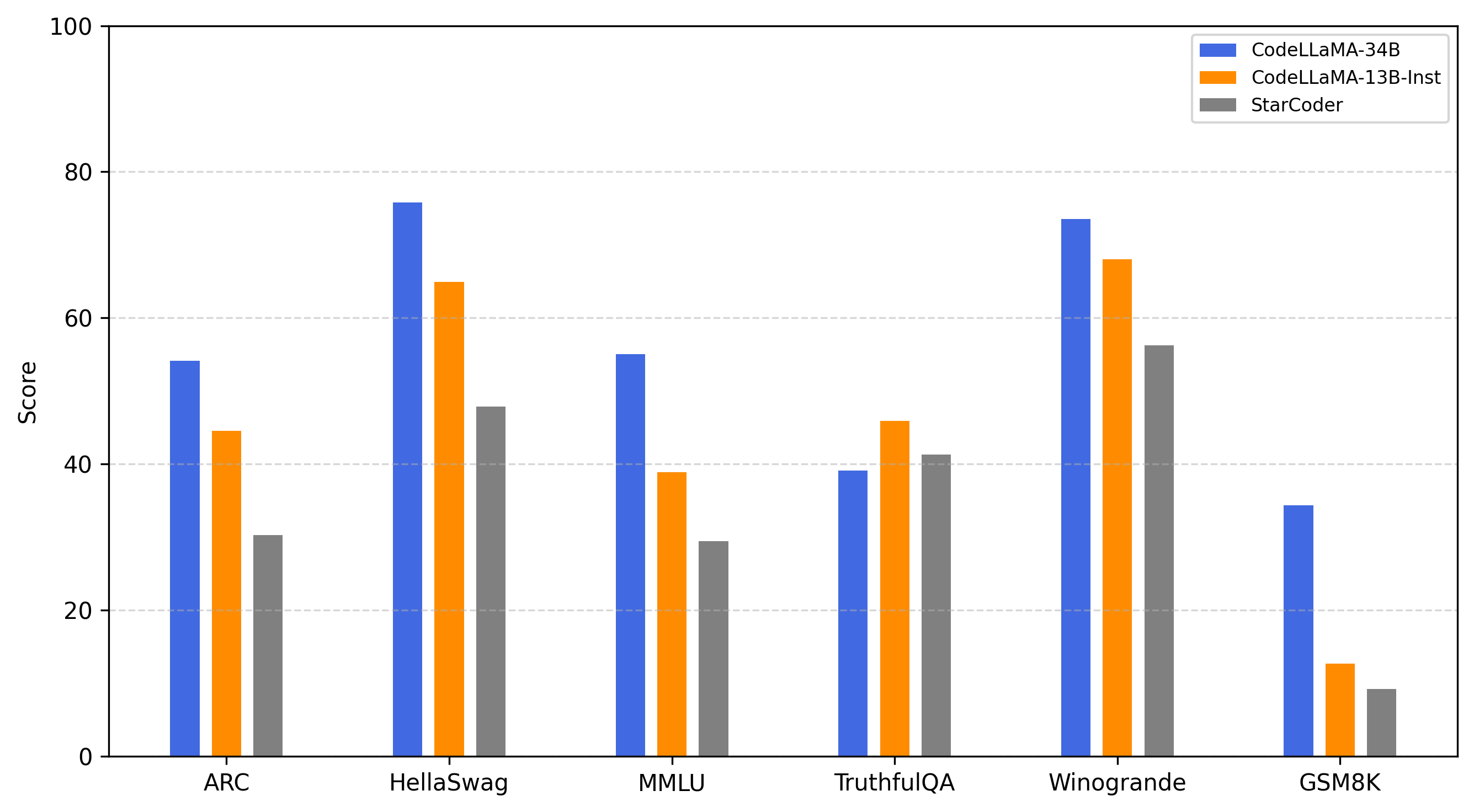}
\label{fig:code_llm_metrics_nl}
\end{subfigure}
\vspace{-8mm}

\begin{subfigure}{0.49\textwidth}
\centering
\caption{Code specific LLMs on Code Explanation metrics}
\includegraphics[width=\textwidth]{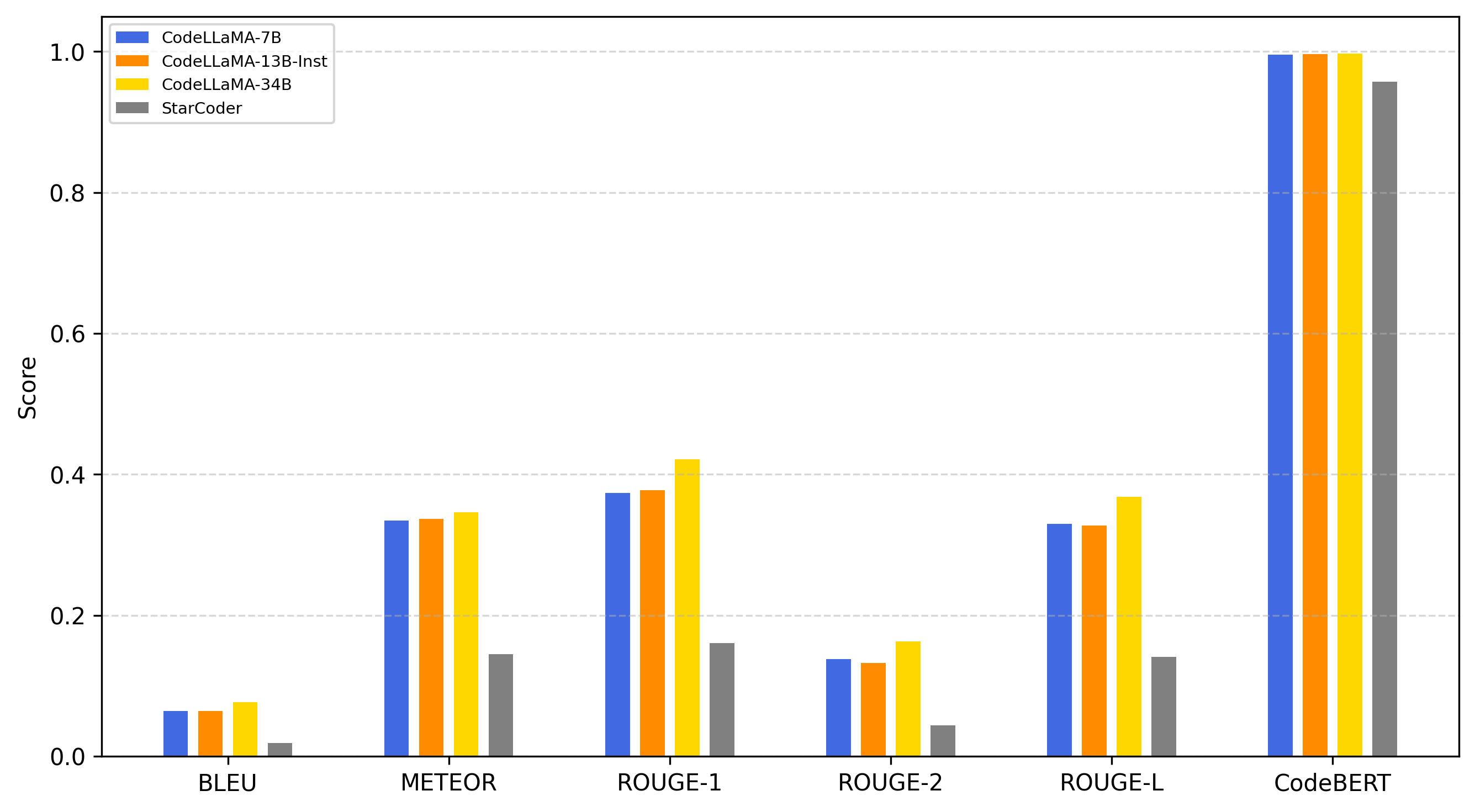}
\label{fig:code_llm_metrics_code}
\end{subfigure}
\vspace{-8mm}
\vspace{-5mm}
\end{figure}

\section{Results and Analysis}
In this section we analyse the performance of general purpose and code specific LLMs. The evaluation numbers across different benchmarks are in Figure \ref{fig:combined_metrics}.

\subsection{{Evaluation of General Purpose LLMs}}
In this section we present a comparative analysis of the general purpose LLMs -- LLaMA 2 series (13B and 7B), LLaMA 3-8B, Vicuña-7B, and Mistral-7B, on the NLP and Code Explanation benchmarks.

\subsubsection{Natural Language Metrics}

As shown in Figures~\ref{fig:gen_llm_metrics} , the plot compares the performance of various language models across six different NLP benchmarks: ARC, HellaSwag, MMLU, TruthfulQA, Winogrande, and GSM8K. The models evaluated include Meta's LLaMA 2 series (13B and 7B), LLaMA 3-8B, Vicuña-7B, and Mistral-7B. We could not find the performance metrics for DeepSeek-R1:1.5B in the HuggingFace leaderboard or the paper. Therefore we could not provide the evaluation here. 

\vspace{2mm}
\noindent
\textbf{Model Size: } Larger models like LLaMA-2-13B consistently outperform smaller counterparts, even Llama-3-8B in benchmarks, especially in benchmarks like ARC and MMLU, indicating that scale remains a significant factor in improving model performance. 

\vspace{2mm}
\noindent
\textbf{Performance on NL metrics}: We now discuss the performances of the models across the different linguistic benchmarks.

$\bullet$\textbf{ Linguistic Capabilities}: In the MMLU benchmark, that offers a variety of linguistic tasks, we find \textit{Llama-3-8B} to perform the best across the five LLMs. This is because the model has a larger number of parameters and layers, sophisticated training mechanisms and an extended training dataset.

$\bullet$\textbf{ Commonsense Reasoning}: Benchmark datasets that judge the performance of the LLMs include ARC, HellaSwag and Winogrande. We find that \textit{Llama-2-13B} and \textit{Llama-3-8B} models outperform the remaining models, suggesting their superiority in reasoning tasks. 

$\bullet$\textbf{ Mathematical Reasoning}: In the GSM8K benchmark that assess a model's mathematical problem-solving capabilities, we find \textit{Llama-3-8B} to be the clear winner. It is likely to benefit from the expanded dataset containing a broader range of mathematical problems and concepts and CoT prompting techniques for the task. However, the performance of all the LLMs are poor at this benchmark compared to others.

$\bullet$\textbf{Trustworthiness}: For the TruthfulQA benchmark, we find \textit{Vicuna-7B} to be performing the best, leaving behind Llama-3-8B significantly behind. It has been observed in ~\cite{lin2022truthfulqa} that larger models are less truthful because they are better at learning from the training distribution.

\subsubsection{Code Explanation metrics}: As seen in Figure ~\ref{fig:gen_llm_metrics_code}, for the CoNaLa benchmark, which evaluates a model's ability to interpret and express code snippets in natural language,\textit{ Mistral-7B} stands out as the top performer. \textit{LLaMA3-8B} follows closely, offering a well-balanced performance with strong structural and keyword-level matching. \textit{Vicuna-7B} demonstrates good semantic comprehension and moderate lexical flexibility. \textit{LLaMA2-7B} exhibits strong semantic intent recognition but struggles with surface-level phrasing, while \textit{LLaMA2-13B} performs the weakest overall, lacking in both semantic and lexical alignment.

\subsection{Evaluation of Code specific LLMs}
\label{sec:analysis_cpl}
In this section we present a comparative analysis of the code specific LLMs -- CodeLlama-34B, CodeLlama-13B-Instruct, and Starcoder, on the NLP and Code Explanation benchmarks.

\subsubsection{Natural Language Metrics}
As shown in Figure \ref{fig:code_llm_metrics_nl}, the plot compares the performance of three code-focused language models: CodeLlama-34B, CodeLlama-13B-Instruct, and Starcoder across six NLP benchmarks: ARC, HellaSwag, MMLU, TruthfulQA, Winogrande, and GSM8K. The key observations here as follows:

\vspace{2mm}
\noindent
\textbf{Model Size:} CodeLlama-34B consistently outperforms both CodeLlama-13B-Instruct and Starcoder-15.5B across all benchmarks, with the most significant margin observed in the HellaSwag benchmark. This suggests that model size significantly contributes to improved performance, especially in tasks involving common sense and natural language understanding.

\vspace{2mm}
\noindent
\textbf{Performance on NL metrics}:  We now discuss the performances of the models across the different linguistic benchmarks.

$\bullet$ \textbf{ Linguistic Capabilities:} We find  \textit{CodeLlama-34B} model to have the best performance in the MMLU benchmark. This suggests that the model is robust in handling both general NLP and code-related tasks, making it a versatile model in the domain.

$\bullet$ \textbf{ Commonsense Reasoning:} Again we find \textit{CodeLlama-34B} to be outperforming the other code-specific LLMs. Specifically, for ARC and HellaSwag, we find that the differences are notably higher compared to Winogrande. This indicates that certain benchmarks are more sensitive to the model’s scale and possibly its training data.

$\bullet$ \textbf{ Mathematical Reasoning:} On this benchmark, we find \textit{CodeLlama-34B} to be performing the best. As a general observation, the performance of the Code-LLMs are far less in GSM8K compared to others. This suggests the inherent difficulty of the benchmark and opens up several research avenues. The cross-domain analysis reveals that code specific LLMs in general fail at mathematical reasoning. We believe that the LLMs must possess sufficient mathematical reasoning abilities in order to excel at coding tasks.

$\bullet$ \textbf{ Trustworthiness:} Similar to what was noted in previous section we find the smaller model \textit{CodeLlama-13B} to be performing the best. 

\subsubsection{Code Explanation metrics}: For the CoNaLa benchmark, which assesses code-to-language alignment, as we find in Figure~\ref{fig:code_llm_metrics_code} \textit{CodeLLaMA-34B} is the best-performing model, consistently achieving the highest scores across all evaluation metrics (BLEU, METEOR, ROUGE, CodeBERT). It excels at generating fluent and semantically precise queries that closely resemble human annotations. Both \textit{CodeLLaMA-7B} and \textit{CodeLLaMA-13B-Instruct} show comparable performance, with the instruct variant slightly outperforming in semantic intent understanding. \textit{StarCoder} significantly underperforms compared to the CodeLLaMA models, displaying weaker lexical and semantic alignment. \textit{DeepSeek-R1:1.5B} ranks the lowest, indicating limited ability in understanding and verbalizing code intent effectively.\\\\
\textbf{Effect of Instruction Tuning:} CodeLlama-13B-Instruct generally performs better than Starcoder-15.5B but falls short compared to CodeLlama-34B counterpart. This suggests, that for a bigger model instruction tuning in enhances model performance. However, for a smaller model training from scratch might be more effective~\cite{ouyang2022training, allen2024physics}.

\begin{table*}[!thb]
\caption{Summary of benchmarks with best-performing general-purpose and code-specific LLMs along with performance differences. (*) indicates a statistically significant difference between  general-purpose and code-specific models (p-value \textless{} 0.05).}
\label{tab:combined-table-horizontal}
\centering
\resizebox{\textwidth}{!}{ 
\begin{tabular}{|c|c|c|c|c|}
\hline
\textbf{Benchmark Type} & \textbf{Benchmark Name} & \textbf{Best General-Purpose LLMs} & \textbf{Best Code-Specific LLMs} & \textbf{Performance Difference} \\
\hline
Linguistic & MMLU & Llama-3-8B & CodeLlama-34B & 11.67 \\
\hline
Commonsense Reasoning & ARC & Mistral-7B & CodeLlama-34B & 5.88 \\
\hline
Commonsense Reasoning & HellaSwag & Mistral-7B & CodeLlama-34B & 7.49 \\
\hline
Commonsense Reasoning & Winogrande & Mistral-7B & CodeLlama-34B & 4.81 \\
\hline
Mathematical Reasoning & GSM8K & Llama-3-8B & CodeLlama-34B & 11.45 \\
\hline
Trustworthiness & TruthfulQA & Vicuna-7B & CodeLlama-13B & 4.53 \\
\hline
\multirow{6}{*}{Code Explanation (CoNaLa)} & BLEU & Mistral-7B & CodeLLaMA-34B & 0.0195\textsuperscript{*} \\
\cline{2-5}
& METEOR & Mistral-7B & CodeLLaMA-34B & 0.0016 \\
\cline{2-5}
& ROUGE-1 & Mistral-7B & CodeLLaMA-34B & 0.0352\textsuperscript{*} \\
\cline{2-5}
& ROUGE-2 & Mistral-7B & CodeLLaMA-34B & 0.0316\textsuperscript{*} \\
\cline{2-5}
& ROUGE-L & Mistral-7B & CodeLLaMA-34B & 0.0354\textsuperscript{*} \\
\cline{2-5}
& CodeBERTScore & Vicuna-7B & CodeLLaMA-34B & 0.0015\textsuperscript{*} \\
\hline
\end{tabular}
}
\end{table*}
\subsection{Cross-Domain Performance}
\label{sec:crossdomain}

In real-world scenarios, a single use-case may span across multiple domains, requiring an LLM to demonstrate robust cross-domain capabilities. This section highlights which models generalize best across both natural language and code-related tasks by combining insights from the benchmark results.

Table~\ref{tab:combined-table-horizontal} compares the performance of general-purpose LLMs (e.g., Llama-3-8B, Mistral-7B, Vicuna-7B) with code-specific LLMs (e.g., CodeLlama-13B, CodeLlama-34B) across diverse reasoning and language understanding benchmarks. The last column shows the absolute  difference in performance between the best general purpose LLM and the best code specific LLM for the corresponding benchmark.

Across linguistic (MMLU) and reasoning benchmarks (ARC, HellaSwag, Winogrande, GSM8K), \textit{CodeLlama-34B} consistently outperforms the strongest general-purpose counterparts, with performance differences ranging from 4.8 to 11.7 points. The largest margins are observed for MMLU (+11.67) and GSM8K (+11.45), suggesting that the structured, syntax-aware training of code models enhances both factual reasoning and mathematical problem-solving. Even in commonsense reasoning tasks, where general-purpose models are typically strong, \textit{CodeLlama-34B} maintains a measurable advantage ($\sim$ 5–7 points), implying a positive transfer of code-related logical representations to natural-language reasoning.

In trustworthiness evaluation (TruthfulQA), \textit{CodeLlama-13B} exceeds Vicuna-7B by 4.53 points, showing that code-tuned models may exhibit more factually grounded or deterministic generation behavior-potentially due to their exposure to structured, verifiable information during training.

For the Code Explanation (CoNaLa) task, the fine-grained text-generation metrics (BLEU, METEOR, ROUGE, CodeBERTScore) also favor \textit{CodeLlama-34B}. Although the absolute differences are numerically smaller ($\sim$0.001–0.035), most are statistically significant (p \textless{} 0.05), confirming that the improvement is consistent and meaningful.

Overall, the results indicate that code-specific LLMs may generalize beyond programming contexts, leveraging their disciplined training on structured data to improve reasoning, mathematical accuracy, and textual precision. Meanwhile, general-purpose models like Mistral-7B and Llama-3-8B remain competitive but exhibit comparatively weaker logical and compositional consistency. This pattern underscores that code-centric pretraining enhances cross-domain reasoning capabilities and that even for non-coding tasks can show measurable performance gains.



\subsection{Application to Model Selection}
Model selection is a critical component of any machine learning pipeline and extends naturally to the process of choosing suitable Large Language Models (LLMs) for downstream applications. The comparative results discussed in the previous section and summarized in Table~\ref{tab:combined-table-horizontal} can directly inform such decisions by aligning model characteristics with task-specific requirements. Below, we outline several representative scenarios that illustrate how empirical performance differences can guide practical model selection.

\vspace{2mm}
\noindent
\textbf{General NLP use-case emphasizing trustworthiness:}
For applications that demand reliable and factually consistent language generation-such as automated summarization, report generation, or conversational agents where factual accuracy outweighs linguistic fluency-the results indicate that \textbf{Vicuna-7B} offers the best trade-off. Although it may not achieve the highest linguistic benchmark scores, its strong performance on TruthfulQA suggests higher reliability and reduced susceptibility to hallucination.

\vspace{2mm}
\noindent
\textbf{Mathematical reasoning under memory constraints:}  
Tasks involving numerical reasoning, structured problem-solving, or mathematical question answering-while operating under limited computational resources-benefit from models that combine reasoning ability with efficiency. In this setting, \textbf{CodeLLaMA-13B} provides an optimal balance, demonstrating strong performance on GSM8K while maintaining a moderate parameter size suitable for resource-limited environments.

\vspace{2mm}
\noindent
\textbf{Code-to-text generation with high lexical and syntactic precision:}  
For scenarios such as code summarization, documentation generation, or intent extraction from Python code, where precision and syntactic fidelity are paramount, \textbf{CodeLLaMA-34B} emerges as the most suitable model. Its consistent superiority across BLEU, ROUGE, and CodeBERTScore metrics in the CoNaLa benchmark highlights its capability to produce semantically faithful and linguistically coherent code explanations.

\vspace{2mm}
\noindent
\textbf{Lightweight semantic explanation of code snippets:}  
In educational or assistive contexts-such as automated code commenting, interactive learning tools, or embedded IDE assistants-efficiency and responsiveness often take precedence over maximal accuracy. Here, \textbf{Mistral-7B} stands out as the most efficient general-purpose LLM, offering competitive performance across multiple reasoning tasks while remaining significantly lighter than larger models like LLaMA-3-70B.

In summary, this analysis underscores that no single LLM is universally optimal; rather, the suitability of a model depends on the interplay between task characteristics (e.g., reasoning vs. linguistic precision), resource availability, and desired output reliability. Empirical benchmarks such as those presented here provide an evidence-based foundation for informed model selection, allowing practitioners to align model choice with specific performance and efficiency trade-offs.

\section{Conclusion and Future Work}
This work presented a unified evaluation of general-purpose and code-specific Large Language Models (LLMs) across benchmarks assessing linguistic competence, reasoning, and trustworthiness. Our findings show that code-centric models, such as CodeLLaMA and StarCoder, generalize effectively beyond programming contexts, demonstrating stronger reasoning and compositional consistency. In contrast, general-purpose models like Mistral-7B and Llama-3-8B excel in linguistic fluency but show weaker logical coherence. These results suggest that structured code pretraining imparts inductive biases beneficial for cross-domain reasoning. The analysis also emphasizes the value of informed model selection, providing practical guidance for aligning LLM choice with task-specific needs. 

However, the scope of this study remains bounded by several factors. We examined a limited subset of LLMs  amid a rapidly expanding ecosystem. Future work should incorporate a broader and more diverse set of models to enhance the generalizability of these findings. Additionally, while our analysis focused primarily on natural language generation benchmarks, a deeper exploration of code-centric tasks such as code generation, test case generation, and debugging assistance would provide a more comprehensive understanding of model capabilities. Expanding evaluations across both model families and task types will be key to driving future advancements in LLM research and application.

\bibliographystyle{ACM-Reference-Format}
\bibliography{references}










\end{document}